# An Observation of a Circular Motion using Ordinary Appliances: Train Toy, Digital Camera, and Android based Smartphone

S. Viridi*, T. Moghrabi, and M. Nasri

*Abstract*

*Using a digital camera (Sony DSC-S75) in its video mode and a smartphone (Samsung GT-N700) equipped with an acceleration sensor, observation of a uniform circular motion of a toy train (Thomas & Friends, Player World, CCF No. 2277-13) is conducted. From the first observation average centripetal acceleration about 0.154 m/s$^2$ is obtained, while the second gives 0.350 m/s$^2$ of average centripetal acceleration by assuming ideal condition, where measured accelerations in z direction is not interpreted.*

**Introduction**

Uniform linier motion (ULM), non-uniform linear motion (NLM), simple harmonic motion (SHM), and Uniform circular motion (UCM) are examples of simple motion that is usually taught in teaching motion in physics and they are related to Newton equations of motion (NEM) [1]. ULM, NLM, and SHM can be derived smoothly from NEM but not UCM, since the last requires finding solution of two second order differential equations that are coupled [2]. Other problem with UCM is when observation taken place at the moving object, since its frame of reference (FR) is not inertial to laboratory FR, that means a fiction force must be applied in order to maintain the validity of NEM [3]. Current information technology innovation in a form of smartphone equipped with acceleration sensor opens an opportunity to measure acceleration of moving object with only a few difficulties [4]. Other way to observe motion of an object is by analyzing its recorded movie [5, 6]. Observation of a UCM using a digital camera in video mode and a smartphone with its acceleration sensor is reported in this work.

**Theoretical background**

Let us define two FR that are labeled with $O$ and $O'$, where $O$ stands for laboratory FR and $O'$ for moving object FR. Along with $O$ there are $x$, $y$, and $z$ coordinates and also with $O'$ there are $x'$, $y'$, and $z'$. For a 2-d UCM that moves only in $xy$ plane with constant angular velocity $\omega$, origin of moving object FR can be written as

$$X = (X_0 - R\cos\varphi_0) + R\cos(\omega t + \varphi_0), \tag{1}$$

$$Y = (Y_0 - R\sin\varphi_0) + R\sin(\omega t + \varphi_0), \tag{2}$$

where $X = X_0$ and $Y = Y_0$ at $t = 0$. Actually, moving object RF in this case is not easily defined since it rotates while the object moves as it is illustrated in Figure 1. Position of O' is simply given by Equation (1) and (2), but the orientation of $x'$ and $y'$ depend on value of $\omega t$.

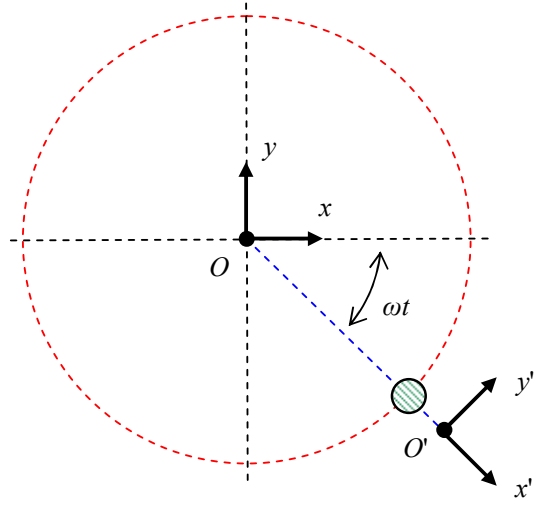

Figure 1. Two different RF: O for laboratory and O' for moving object.

Using relation between RFs, it can be written that

$$x = X + x' = (X_0 - R\cos\varphi_0) + R\cos(\omega t + \varphi_0) + x', \tag{3}$$

$$y = Y + y' = (Y_0 - R\cos\varphi_0) + R\cos(\omega t + \varphi_0) + y'. \tag{4}$$

If the moving object is placed on the origin ($x' = 0$, $y' = 0$) and is always at rest according to O', then $x$ and $y$ from Equations (3) and (4) will be equal to Equations (1) and (2).

Until now, the force that maintains the object to moving with O' has not yet been discussed. If there is no net force according to O, then the object can not stick and then moves together with O'. On the other side, there should be no net force according to O' since the object is always at rest. Suppose that there are net forces $F_x'$ and $F_y'$ according to O' (they will be later proved to be zero), then NEM in O' are

$$m\frac{d^2 x'}{dt^2} = F_x', \tag{5}$$

$$m\frac{d^2 y'}{dt^2} = F_y'. \tag{6}$$

According to $O$ there must be also net forces, with similar form to Equation (5) and (6). Deriving Equations (3) and (4) two times with respect to time $t$ and then substitute Equation (5) and (6) and also for $F_x$ and $F_y$, will produce

$$\frac{F_x'}{m} = \frac{F_x}{m} + \omega^2 R \cos(\omega t + \varphi_0), \qquad (7)$$

$$\frac{F_y'}{m} = \frac{F_y}{m} + \omega^2 R \sin(\omega t + \varphi_0). \qquad (8)$$

The terms $F_x'/m$ and $F_y'/m$ will be measured by acceleration sensor while average ω will be obtained from recorded observation video.

An object that performs a circular motion with tangential velocity $v_T$ at radius $R$ will give angular velocity

$$\omega = \frac{v_T}{R} \qquad (9)$$

and centripetal acceleration

$$a_C = \omega^2 R. \qquad (10)$$

Tangential velocity $v_T$ is calculated for every quarter of circle circumference

$$v_T = \frac{s(t + \Delta t) - s(t)}{\Delta t}, \qquad (11)$$

where for ideal condition, the relation

$$v_T = \frac{\pi R}{2 \Delta t} \qquad (12)$$

will hold with $\Delta t$ is the time needed for the moving object to elapse length of a quarter of circle circumference.

**Experimental setup**

Train toy (Thomas & Friends, Player World, CCF No. 2277-13) has track radius $r = 36$ cm and track width $\Delta r = 3$ cm. Motion of the train is recorded from above using a digital camera (Sony DSC-S75) with 160 px × 112 px video size and 25 fps. The video is then analyzed using Media Player Classic (version 6.4.9.1 rev 86) with $\Delta t$ = 0.5 s accuracy. A smartphone (Samsung GT-N700) is equipped K3DH acceleration sensor and acquisition software SensorLogger (10 ms sampling rate) and Sensor Ex (100 ms sampling rate). The smartphone is attached on top of the train toy and it recorded the acceleration as the train performed a circular motion.

**Results and discussion**

Observation of train toy motion using digital camera in its video mode will produce a series of images, where four of them are given as illustration in Figure 2. The times when the train toy arrives at every quarter of circle circumference are recorded and these data are listed in Table 1. Values in most right column in the table are obtained from the other two columns using Equation (12) by assuming an ideal condition from Equation (11). Average tangential velocity $v_T$ is found about 0.235 m/s, which produces 0.154 m/s$^2$ for centripetal acceleration through Equations (9) and (10).

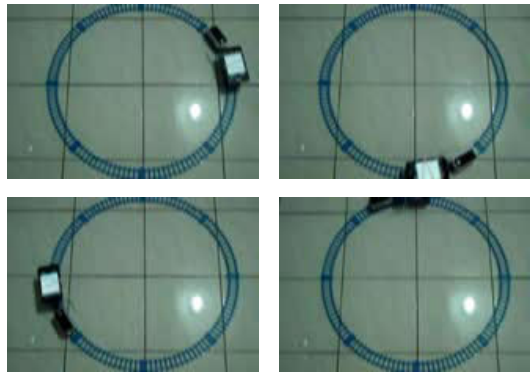

Figure 2. Train toy performs a clockwise circular motion at time *t*: (a) 13 s, (b) 15 s, (c) 18 s, and (d) 20 s.

Acceleration sensor in the smartphone attached on the train toy produces time series for acceleration in *x*, *y*, and *z* direction as given in Figure 3. Two different softwares, which are SensorLogger and Sensor Ex, are used for obtaining acceleration data. They seem to produce similar data as illustrated in Figure 3. Measured acceleration in *z* direction must be first subtracted with gravitation acceleration *g* since the sensor also measures *g*. Even acceleration data in this direction is not so important but it is a good sign that the sensor reports reading values that have physical meaning. Averaged values for $a_x'$, $a_y'$, and $a_z'$-$g$ are -0.336 m/s$^2$, -0.097 m/s$^2$, and -0.481 m/s$^2$, respectively.

For ideal condition, average of Equations (7) and (8) will give acceleration in *x* and *y* direction in lab FR, respectively. By assuming this condition for data in Figure 3 (top) and (middle), value of total acceleration 0.350 m/s$^2$ is obtained according to lab FR, which consist only the centripetal acceleration.

Table 1. Observed elapsed distance $s$ from recorded video and calculate tangential velocity $v_T$ for time $t$.

| $t$ (s) | $s$ (m) | $v_T$ ( m/s) |
|---|---|---|
| 0 | 0.000 | 0.188 |
| 3 | 0.565 | 0.283 |
| 5 | 1.131 | 0.188 |
| 8 | 1.696 | 0.283 |
| 10 | 2.262 | 0.188 |
| 13 | 2.827 | 0.283 |
| 15 | 3.393 | 0.188 |
| 18 | 3.958 | 0.283 |
| 20 | 4.524 | 0.188 |
| 23 | 5.089 | 0.283 |
| 25 | 5.655 | 0.188 |
| 28 | 6.220 | 0.283 |
| 30 | 6.786 | 0.188 |
| 33 | 7.351 | 0.283 |
| 35 | 7.917 | 0.188 |
| 38 | 8.482 | 0.283 |
| 40 | 9.048 | 0.188 |
| 43 | 9.613 | 0.283 |
| 45 | 10.179 | 0.188 |
| 48 | 10.744 | 0.283 |
| 50 | 11.310 | 0.188 |
| 53 | 11.875 | 0.283 |
| 55 | 12.441 | 0.188 |
| 58 | 13.006 | 0.283 |
| 60 | 13.572 | 0.226 |

Both values of obtained centripetal acceleration $0.154$ m/s$^2$ and $0.350$ m/s$^2$ can not be guaranteed too much, since they are much smaller than the fluctuated measured values in Figure 3 (top) and (middle). More improvement is needed to overcome this problem. One of the suggestions is by driving the toy train with higher tangential velocity to produce higher centripetal force compared to the fluctuation.

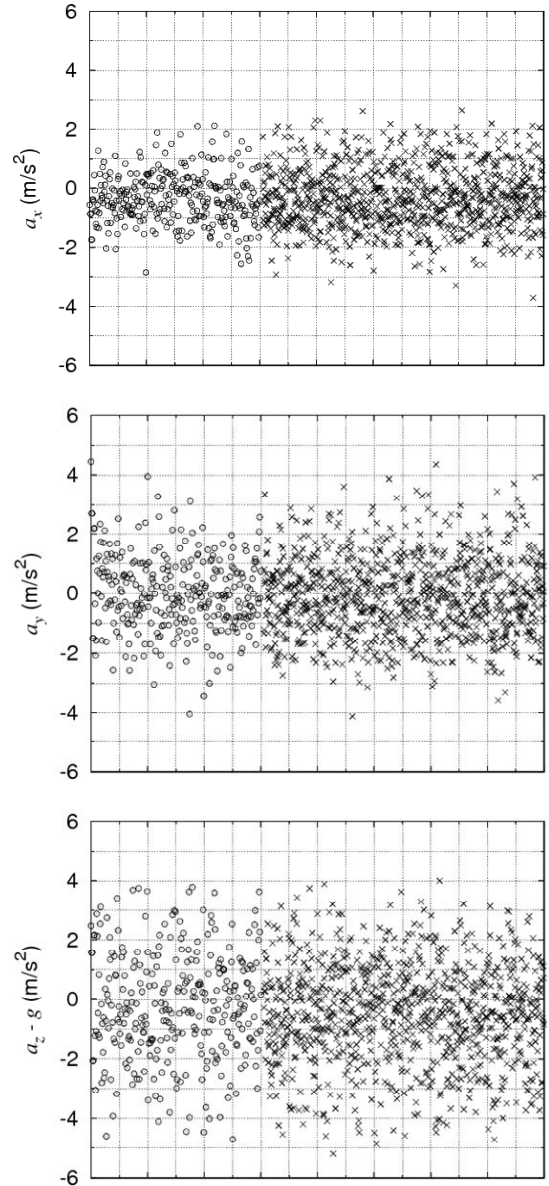

Figure 3. Acceleration measured by K3DH acceleration sensor for: $x'$ (top), $y'$ (middle), and $z'$ (bottom) direction using SensorLogger (O) and Sensor Ex (×) software in moving object FR.

**Conclusion**

Observations of train toy performing a circular motion have been conducted using video and acceleration sensor. From both observations different values of averaged centripetal acceleration are obtained, which are $0.154$ m/s$^2$ and $0.350$ m/s$^2$. These values are still too small compare to fluctuation observed in the acceleration sensor, which means both values are still questionable to be trusted. Improvement is needed to overcome this problem.

**Acknowledgments**

SV would like to thank Prof. Mikrajuddin Abdullah, Dr. Suprijadi, and Mr. Agus Suroso for fruitful discussion atmosphere in analyzing the data and also for research program Riset Inovasi KK ITB (RIK-ITB) in year 2013 with contract number 248/I.1.C01/PL/2013 for supporting this work.

Sparisoma Viridi*
Nuclear Physics and Biophysis Research Division
Institut Teknologi Bandung
dudung@gmail.com

Tarex Moghrabi
Mechatronics Engineer, Sensor Ex, TarCo,
URI sites.google.com/site/tarsensorics
tarexmo@gmail.com

Meldawati Nasri
Sukma Bangsa Bireun High School, Bireun,
Aceh 24251, Indonesia
cimemeloge@yahoo.com

*Corresponding author